\documentclass[fleqn,12pt,twoside]{article}
\usepackage[headings]{espcrc1}

\readRCS
$Id: espcrc1.tex,v 1.2 2004/02/24 11:22:11 spepping Exp $
\usepackage{epsfig}
\usepackage[figuresright]{rotating}

\newcommand{\beq}{\begin{equation}}
\newcommand{\eeq}{\end{equation}}
\def\bea{\begin{eqnarray}}  \def\eea{\end{eqnarray}}

\def\simge{\mathrel{%
   \rlap{\raise 0.511ex \hbox{$>$}}{\lower 0.511ex \hbox{$\sim$}}}}
\def\simle{\mathrel{
   \rlap{\raise 0.511ex \hbox{$<$}}{\lower 0.511ex \hbox{$\sim$}}}}

\newcommand{\be}{\begin{equation}}
\newcommand{\ee}{\end{equation}}

\newcommand{\AmS}{{\protect\the\textfont2
  A\kern-.1667em\lower.5ex\hbox{M}\kern-.125emS}}

\title{Color strings, pomeron shadowing and Color Glass Condensate}

\author{E. G. Ferreiro\address[USC]{Departamento de F\'{\i}sica
        de Part\'{\i}culas,\\ Universidad de Santiago de Compostela,\\
        15782 Santiago de Compostela, Spain}}

\runtitle{Color strings, pomeron shadowing and CGC}
\runauthor{E. G. Ferreiro}

\begin{document}

\maketitle

\begin{abstract}
I review the problem of parton saturation and its implications through three in principal 
different
approaches, but somewhat related: saturation in a geometrical approach, QCD saturation through the
Color Glass Condensate and shadowing in a Pomeron approach.
\end{abstract}
\vskip 1.0truecm


\section{Introduction}
In the recent experiments like DIS at HERA or the heavy-ion experiments at RHIC, and
also in expected LHC at CERN, the number of involved partons is very large,
due to the high energy and/or the high number of participants of those experiments.
These high parton densities should in principal lead
to an extremely huge multiparticle production, but experimentally we have seen that
this is not the case. So there should be a mechanism that reduces the number of created
particle. Two kind of phenomena have been proposed. On one side there is the possibility
of saturation in the initial state of the collision. On the othe side, there is the proposal
of the creation of a Quark Gluon Plasma,
and it has been presented as a final state interaction mechanism.
Here, I review the problem of parton saturation and its implications through three in principal
different
approaches, but somewhat related: saturation in a geometrical approach, QCD saturation through the
Color Glass Condensate and shadowing in a Pomeron approach.

\section{Geometrical approach to saturation}
\subsection{String models and percolation}

In many models of hadronic and nuclear collisions, color strings are exchanged between 
the projectile and the target. Those strings act as {\it color sources} of particles through the creation
of $q-{\bar q}$ pairs from the sea. 
The number of strings grows with the energy and with the number of nucleons of the 
participant nuclei. 


In impact parameter space these strings are seen as circles inside the total collision area.
When the density of strings becomes high the string color fields begin to overlap and 
eventually individual strings may fuse, 
forming a new string --{\it cluster}-- which has a higher color charge at its ends, 
corresponding to the summation of the color charges located at the ends of the original strings. 
The new string clusters break into hadrons according to their higher color. 
As a result, there is a reduction of the 
total multiplicity. 
Also, as the energy-momenta of the original strings are summed to obtain the energy-momentum of 
the resulting cluster, the mean transverse momentum of the particles created by those clusters
is increased compared to the one of the particles 
created from individual sources.

\begin{figure}
\centering\leavevmode
\epsfxsize=13cm\epsfysize=3.5cm\epsffile{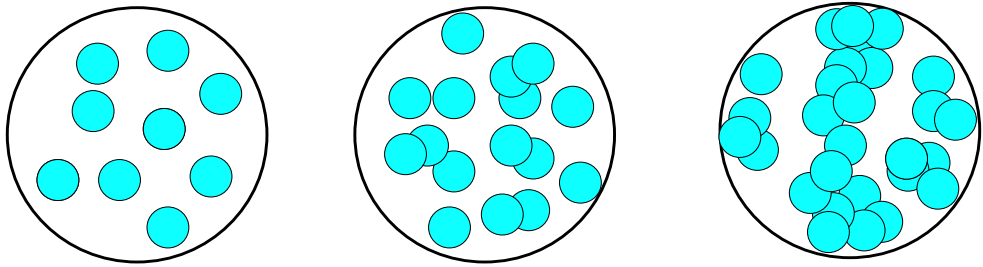}
\caption{From left to righ:
Density of strings in the transverse space, from low energy and/or
low number of participants to high energies and/or high number of
participants. In the last circle we show percolation.}
\label{fig1}
\end{figure}

As the number of strings increases, 
more strings overlap. 
Some years ago, it has been proposed in Ref. \cite{REF96} that
above a critical density of strings {\it percolation}
 occurs, so that paths of overlapping circles are formed through the whole collision area,
as it is represented in Fig. \ref{fig1}.
Along these paths the medium behaves like a color conductor. 
Also in \cite{REF96}, we have made the remark that 
several fused strings can be considered as a domain of a {\it non thermalized Quark Gluon Plasma}. 
The percolation gives rise to the formation of a non thermalized 
Quark Gluon Plasma on a nuclear scale. 

Note that here we are not speaking about a final state interaction phenomenon, since there is no 
thermalization involved. In fact, what we are trying is
to determine under what conditions the initial state 
configurations can lead to color connection, and more specifically, 
if variations of the initial state can lead to a transition from 
disconnected to connected color clusters. 
The results of such a study of the pre-equilibrium state in nuclear collisions 
do not depend on the subsequent evolution and thus in particular not require any kind of 
thermalization.

The structural problem underlying the transition from disconnected to connected systems 
of many components is a very general one, ranging from clustering in spin systems to 
the formation of galaxies. 
The formalism is given by percolation theory, which describes geometric critical behavior.

\subsection{Percolation theory}

Consider placing $N$ small circular discs (color sources, strings or partons) 
of radius $r$ onto a large circular manifold (the transverse nuclear plane) of radius $R$; 
the small discs may overlap. 
With increasing density 
\be
\eta = \frac{N \pi r^2}{ \pi R^2}\  , 
\label{ec1}
\ee
this overlap will lead to 
more and larger connected clusters. 
The most interesting feature of this phenomenon is that the average cluster size 
increases very suddenly from very small to very large values. 
This suggests some kind of geometric critical behavior. In fact, 
the cluster size diverges at a critical threshold value $\eta_c$ of the density. This appearance
of an infinite cluster at $\eta= \eta_c$ is defined as percolation: the size of the cluster 
reaches the size of the system. 
$\eta_c$ has been computed using Monte Carlo simulation, direct connectedness expansion and 
other different methods. All the results are in the range  $\eta_c = 1.12 \div 1.175$.

In our model \cite{REF96} we had proposed a {\it fixed radius} for the independent 
color sources of 
$r=0.2 \div 0.25$ fm,
 that corresponds to a momentum around 1 GeV. 
This value has been obtained from 
Monte Carlo simulations in the framework of the String Fusion Model Code (SFMC) \cite{REFSFMC} 
made at SPS
energies. 
According to eq. (\ref{ec1}), in order to estimate the density $\eta$, one needs to know 
the number of sources $N$. In our model, it is obtained from the SFMC,
that, for nucleus-nucleus collisions, 
takes into account two contributions: one proportional to the number of participant
nucleons --valence-like contribution-- and another one proportional to the number of 
inelastic nucleon-nucleon collisions.
Note that $N$ will depend on the energy $\sqrt{s}$ (or equivalently, on $x$) and on the number of 
participant nucleons $A$, so in some way the condition to achieve percolation depends on 
$A$ and $s$, $\eta=\eta(A,x)$.
$\pi R^2$ corresponds simply to the nuclear overlap area, $S_A$, 
at the given impact parameter. This overlap area can be determined in 
a Glauber study, using Woods-Saxon nuclear profiles.
That leads to the following results: In our model,
at SPS energies, the critical threshold for percolation could eventually been 
achieved for the most central Pb-Pb collisions, and for sure in Au-Au central collisions at 
RHIC energies and even in p-p collisions at LHC energies.

\subsection{%
The size of the color sources}

Let us now introduce a new question:
If you look at a fast nucleon coming at you, what do you see? 
It depends on who is looking. Another nucleon sees a disc of radius $r \simeq 1$ 
fm and a certain greyness. 
A hard photon, with a resolution scale $Q^{-1} << 1$ fm, 
sees a swarm of partons. 
How many there are depends on the resolution scale: 
given a finer scale, you can see smaller partons, and there are more the harder you look.
The partons in a nucleon have a transverse size $r_T$ 
determined by their transverse momentum $k_T$ , with $r_T \sim 1/k_T$ . The scale $Q^{-1}$ 
specifies the minimum $k^{-1}_T$ resolved, so the probing photon sees all partons in the range 
$0 \le k_T \le Q$, or equivalently it sees all the partons with a radius $r_T \ge Q^{-1}$.

So the partonic size is through the uncertainty relation determined by its 
average transverse momentum, 
$r^ 2 \sim 1/<k_T^2>$, for a given resolution scale, $<k_T^2> \sim Q^2$.

\begin{figure}
\centering\leavevmode
\epsfxsize=13cm\epsfysize=3.5cm\epsffile{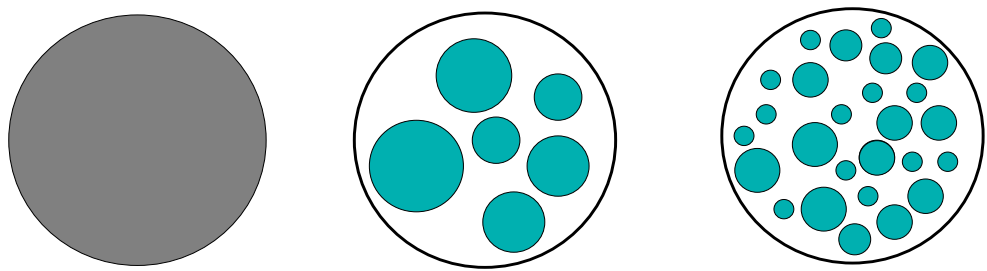}
\caption{From left to righ:
The structure of an incoming nucleon seen for increasing resolution.}
\label{fig2}
\end{figure}

\begin{figure}
\centering\leavevmode
\epsfxsize=12cm\epsfysize=7cm\epsffile{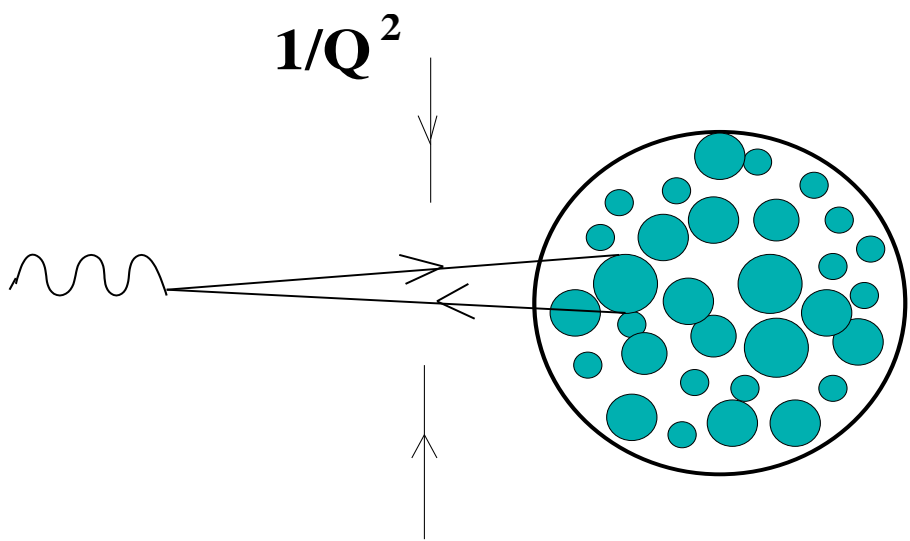}
\caption{
The resolution scale.}
\label{fig3}
\end{figure}

These ideas are illustrated in Figs. \ref{fig2} and \ref{fig3}.

In order to know if the percolation density is achieved in a collision,
we need to compute our eq. (\ref{ec1}). Then it is necessary to know 
the number of initially created sources (partons or strings, it depends on the
kind of model we are using) $N$, and the size of those sources $r$. Remember that in the previous
section we have used a fixed size of $r=0.2 \div 0.25$ fm that corresponds to a moment of 1 GeV.
In \cite{REF96} $N$ is calculated in the framework of the SFMC.

In \cite{REFSATZ}, the number of sources $N$ is calculated through the Wonded Nucleon Model for 
nucleus-nucleus collisions, using P.D.F.'s
for the nucleonic density.
The authors use here an slightly different strategy: Instead of fixing the radius of the initially
created sources, they estimate the momentum of those sources that will lead to percolation.
They transform then eq. (\ref{ec1}) into the following one:
\be
\eta_c = \frac{N \pi r^2_c}{\pi R^2}= \frac{N}{Q_c^2 R^2}\ .  
\label{ec3}
\ee
The condition for percolation, taking $\eta_c=1.12$
will be then:
\be
Q_c^2=\frac{N}{1.12 R^2}\ .
\label{ec4}
\ee
They find that for central Pb-Pb collisions at SPS energies, 
the critical momentum for percolation is $Q^2_c \approx 1$ GeV$^2$, in accordance with our result.
For central Au-Au collisions at RHIC, $Q^2_c \approx 2.5$ GeV$^2$.
Note that again the condition to achieve percolation depends on $A$ and $\sqrt{s}$,
$Q_c=Q_c(A,x)$. 

Beyond the percolation point, one has a condensate, containing 
interacting and hence color-connected sources of all scales $k_T \le Q$. 
The percolation point thus specifies the onset of color deconfinement; 
it says nothing about any subsequent thermalization.

\section{The Color Glass Condensate}

Now we arrive to another approach, the QCD saturation through the formation 
of a Color Glass Condensate (CGC) \cite{REFCGC}. The idea is the following:
At high energy, the QCD cross-sections are controlled by small longitudinal
momentum gluons in the hadron wave function, whose density grows rapidly with
increasing energy or decreasing $x$, due to the enhancement of radiative
process. If one applies perturbation theory to this regime, one finds
that, by resumming dominant radiative corrections at high energy, the BFKL
equation leads to a gluon density that grows like a power of $s$ and in
consequence
to a cross-section that violates the Froissart bound.
Nevertheless,
 the use of perturbation theory to high-energy problems is not obvious. In
fact, the BFKL and DGLAP equations are linear equations that neglet the
interaction among the gluons. With increasing energy, recombination effects
--that are non-linear--
favored by the high density of gluons should become more important and lead
to an eventual {\it saturation} of parton densities.

These effects become important
when the interaction probability for the gluons becomes of order one.
Taking $\frac{\alpha_s N_c}{Q^2}$ as the transverse size of the gluon and
$\frac{x G(x,Q^2)}{\pi R^2}$ as the density of gluons, the interaction
probability is expressed by
\be
\frac{\alpha_s N_c}{Q^2}\,\,\times\,\,
\frac{x G(x,Q^2)}{\pi R^2}\ .
\label{ec5}
\ee
Equivalently, for a given energy, saturation
occurs for those gluons having a sufficiently large transverse size $r_\perp^2
\sim 1/Q^2$, larger than some critical value $1/Q_s(x,A)$. So the phenomenon
of saturation introduces a characteristic momentum scale,
the {\it saturation momentum} $Q_s(x,A)$, which is a measure of the
density of the saturated gluons, and grows rapidly with $1/x$ and
$A$ (the atomic number).
The probability of interaction
--that can be understood as "overlapping" of the gluons in the transverse space--
becomes of order one for those gluons with
momenta $Q^2 \simle Q_s(x,A)$ where
\be
Q^2_s(x,A)=\alpha_s N_c \ \frac{x G(x,Q^2_s)}{\pi R^2} \equiv
\frac{({\rm color\,\, charge})^2}{{\rm area}}\ .
\label{ec6}
\ee

For $Q^2\simle Q^2_s(x,A)$, the non-linear effects
are essential, since they
 are expected to soften the growth of the gluon distribution
with $\tau\equiv \ln(1/x)$.
For a nucleus, $x G_A(x,Q^2_s)\propto A$ and
 $\pi R^2_A\propto A^{2/3}$, so eq.~(\ref{ec6})
predicts
$Q^2_s\propto A^{1/3}$. One can estimate the saturation scale by
inserting the BFKL approximation into
eq.~(\ref{ec6}). This gives
(with $\delta\approx 1/3$ and $\lambda\approx c\bar\alpha_s$ in a
first
approximation):
\be
\label{ec7}
Q^2_s(x,A)\,\,\sim\,\,A^{\delta}\, x^{-\lambda}\,,
\ee
which indicates that an efficient way to create a high-density
environment is to combine large nuclei with moderately small values of $x$,
as it is done at RHIC. In fact the estimated momentum for saturation at RHIC
will be $Q_s= 1 \div 2$ GeV, in accordance with the result of the previous section.

We call the high density gluonic matter at small-$x$ described by this
effective theory a {\it Color Glass Condensate}:
{\it Color} since gluons carry color under $SU(N_c)$;
{\it Glass} since
we have classical coherent fields which are frozen over the typical time scales for 
high-energy scattering, 
but randomly changing over larger time scales. 
So we have a random distribution of time-independent
color charges
which is averaged over in the calculation
of physical observables, in order to have a gauge independent formulation
--in analogy to spin glasses--;
and {\it Condensate} because at saturation the gluon density is of
order $1/\alpha_s$, typical of condensates, so we have a system of saturate
gluons that is a Bose condensate.

\section{Shadowing}
Note that in the string models, like Quark Gluon String Model or Dual Parton Model 
each string corresponds to the exchange of pomerons, and the interaction among the strings 
would correspond to interaction among the pomerons through the triple pomeron vertex.
In particular, in the DPM, 
some kind of saturation is also included through the {\it shadowing} in the 
initial state of the collision \cite{REFSHADOW}. In fact, in absence of shadowing,
the DPM at high energies leads
 to multiplicities that scale with 
the number of binary collisions rather than to a scaling with the number of participants. This is a general property of 
Gribov's Reggeon Field Theory which is known as AGK cancellation, 
analogous to the factorization theorem in perturbative QCD and valid for 
soft collisions in the absence of triple Pomeron diagrams.
Because of this, a dynamical, non-linear shadowing has been included in the DPM. It is determined
in terms of 
the diffractive cross-section. It is controlled by triple pomeron diagrams and
it should lead to 
saturation as $s \rightarrow \infty$. In fact, the shadowing corrections in this approach have a positive
contribution to diffraction and a negative one to the total cross-section.

The shadowing is controlled by the following quantity:
\beq
\label{eq11}
f(x,Q^2)=4\pi
\int _{M^2_{min}}^{M^2_{max}}dM^2 \left. \frac{1}{\sigma_{\gamma^*{\rm nucleon}}}
\frac{d\sigma^{\mathcal{D}}
_{\gamma^*{\rm p}}}{dM^2dt}\right\vert_{t=0}F_A^2(t_{min})
\eeq
which is equivalent to (see below for details):
\beq
\label{eq22}
F(s,y^*)
=4\pi\int_{y_{min}}^{y_{max}}dy\ \frac{1}{\sigma_P(s)}\left.\frac{d\sigma^{PPP}}
{dydt}\right\vert_{t=0}F_A^2(t_{min}),
\eeq
where
\beq
F_A(t_{min})=\int d^2b\ J_0(b\sqrt{-t_{min}})T_A(b),
\label{eq2-1}
\eeq
with $t_{min}=-m_N^2x_P^2$ and $m_N$ the nucleon mass. It represents the coherence effects,
and at RHIC or higher energies can be taken equal to 1.

Quantities as
$\frac{1}{\sigma_{\gamma^*{\rm nucleon}}}
\frac{d\sigma^{\mathcal{D}}_{\gamma^*{\rm p}}}{dM^2dt}\vert_{t=0}$
or
$\frac{1}{\sigma_P(s)}\left.\frac{d\sigma^{PPP}}
{dydt}\right\vert_{t=0}
$
 have been computed
through $F_2(x,Q^2)$ and $F^{(3)}_{2\mathcal{D}}(Q^2,x_P,\beta)$
data on DIS scatterring.

The above expressions correspond to the case with only two scatterings.
In order to include higher order rescatterings it is neccessary to do an unitarization.
Two possible models apply here, eikonal unitarization or Schwimmer unitarization.
Eikonal unitarization will
produce larger shadowing than Schwimmer, as can be expected by comparing the
second non-trivial order in the expansion of both expressions.
Both approaches have been checked with experimental data from DIS and heavy-ion collisions
by Capella et al. and Frackfurt et at.
Experimental data favorize the Schwimmer unitarization, that leads to:
\beq
\label{eq15}
R(A/{\rm nucleon})^{Sch}(b)=\frac{1}{1+A\ f(x,Q^2)\ T_A(b)}\ .
\eeq

Note that
shadowing in nuclei is usually studied through the
ratios of cross sections per
nucleon for
different nuclei. This is the meaning of R(A/{\rm nucleon}) above.

Then the factor for reduction of multiplicities at fixed impact parameter $b$
is:
\beq
\label{eq18}
R_{AB}(b)=\frac{\int d^2s\  R_A(\vec{s})R_B(\vec{b}-\vec{s})}{T_{AB}(b)}\ .
\eeq
$R_{A(B)}(b)$ is given by the r.h.s.
of (\ref{eq15}) multiplied by $T_{A(B)}(b)$ --nuclear profile functions--, $T_{AB}(b)=\int d^2s\  T_A(\vec{s})T_B(\vec{b}-\vec{s})$, and
with $f(x,Q^2)$ expressed in terms of rapidity.

As I said above, the shadowing is controlled by the quantity
$f(x,Q^2)$ (the shadowing grows with $f$).
Their integrations limits are defined as:
$M^2_{min}=
4m_\pi^2
=0.08$ GeV$^2$,  while the upper one is taken from the condition:
\beq
\label{eq5}
x_P=x\left(\frac{M^2+Q^2}{Q^2}\right)\leq x_{Pmax}
\Longrightarrow M^2_{max}= Q^2\left(\frac{x_{Pmax}}{x}-1\right),
\eeq
with $x_{Pmax}=0.1 \div 0.2$.

In order to expressed it
 as a function of the rapidity of the produced particles
$F(s,y^*)$,
 we have
two possibilities:
the first one is using
(\ref{eq11}), but with the integration limits
inspired by the parton model for hard processes:
for projectile $A$ (target $B$),
\beq
\label{eq20}
x_{A(B)}=\frac{m_T}{\sqrt{s}}e^{\pm y^*},
\eeq
with $y^*>0$ for the projectile hemisphere and $y^*<0$ for the target one, and
$m_T=\sqrt{m^2+p_T^2}$ the transverse mass of the produced particle.
On the other hand, we can also compute the reduction factor from the formula
\beq
\label{eq22b}
F(s,y^*)
=4\pi\int_{y_{min}}^{y_{max}}dy\ \frac{1}{\sigma_P(s)}\left.\frac{d\sigma^{PPP}}
{dydt}\right\vert_{t=0}F_A^2(t_{min}),
\eeq
where $\sigma_P(s)$ is the single Pomeron exchange cross section and
$\frac{d\sigma^{PPP}}{dydt}$ the triple Pomeron cross section. 

Using the standard triple Pomeron formula for the latter, we get
\beq
\label{eq23}
\frac{1}{\sigma_P(s)}\left.\frac{d\sigma^{PPP}}{dydt}\right\vert_{t=0}=
C\Delta \exp{(\Delta y)},
\eeq
with $C=\frac
{g_{pp}^{P}(0)r_{PPP}(0)}{4\Delta}$,
$g_{pp}^{P}(0)$ the Pomeron-proton coupling and $r_{PPP}(0)$
the triple Pomeron coupling, both evaluated at $t=0$.
That leads to the following 
values: $C=0.31$ fm$^2$, $\Delta=0.13$, fixed from DIS data. 
The same integration limits used above correspond to:
\beq
\label{eq24}
y^{(A(B))}_{min}=\ln{\left(\frac{s}{M^{2(A(B))}_{max}}\right)} \sim
\ln{\left(\frac{R_A m_N}{\sqrt{3}}\right)}\ \ {\rm and}\ \
y_{max}^{(A(B))}=\frac{1}{2}\ln{\left(\frac{s}{m_T^2}\right)} \mp y^* \ ,
\eeq
where $y^*$ is the center of mass rapidity and $m_T$ the transverse mass of the produced particle.
Finally we get:
\beq
F(s,y^*)=
\int_{y_{min}}^{y_{max}}dy\
\Delta \exp{(\Delta y)}= C [\exp{(\Delta y_{max})} - \exp{(\Delta y_{min})}] \ .
\eeq

\section{Conclusions}

We have compared different models that takes into account saturation in different ways:
from the semi-phenomenological
fusing color sources picture for the soft domain including percolation, 
the QCD saturation through the Color Glass Condensate
and those which
follow from the pomeron approach and that takes into account shadowing
in the initial conditions.

In fact, it seems that the exchanged of
elemental objects, color sources --strings, partons or pomerons--, should lead to a saturation
in the initial conditions when the densities are high enough.
In particular, in which concerns multiplicity dependence on the number of participants $A$, the cited approaches coincide in the results: when going to very high energies, the yield of created particles will be proportional to the number of participant nucleons when the shadowing/saturation effects are taken into account. In abscence of initial effect those yields behave as the number of collisions, so as $A^{4/3}$.

%
%
%


\end{document}